\documentclass[12pt]{iopart}

\usepackage{graphicx}
\usepackage{color}
\begin{document}

\title[RPV supersymmetric contributions 
to the neutron beta decay]{R-parity violating supersymmetric contributions 
to the neutron beta decay}

\author{Nodoka Yamanaka, Toru Sato and Takahiro Kubota}

\address{Department of Physics, Osaka University, Toyonaka, Osaka 560-0043, Japan}
\ead{yamanaka@kern.phys.sci.osaka-u.ac.jp}
\begin{abstract}
We investigate the contribution to the angular correlation
 coefficients of the neutron beta decay
within the R-parity violating (RPV) minimal supersymmetric 
standard model (MSSM). 
The RPV effects contribute to the scalar interaction at the tree level.
The coefficient of the effective scalar interaction of the nucleon 
is determined in terms of the nucleon mass difference. 
On the basis of  the recent update of the analyses of  the
 superallowed Fermi transitions and the recent measurement of 
transverse polarization of the emitted electrons at PSI, we
 deduce  new upper limits  on the 
RPV couplings and compare them with those obtained through pion decay and atomic electric dipole moment (EDM).
We also give a comprehensive analysis of angular correlations sensitive to the RPV interactions.
\end{abstract}

\maketitle

\section{\label{sec:intro}Introduction}

In recent years, the search for new physics ``Beyond  the 
Standard Model'' has been performed in many ways. A particular 
attention is paid to the low energy approach, which consists of
 observing  small discrepancies between   low energy phenomena 
and the Standard Model (SM) \cite{herczeg1, ramsey1, beck, ramsey,
abele1}. 
Among several others,  neutron decay experiments are expected to 
be very promising for new physics search  and are 
actively planned or under operation by many groups at PSI,
 LANSCE, ILL, J-PARC and so forth.
 
On the theoretical side, the Minimal Supersymmetric Standard 
Model (MSSM) \cite{tata} is known to be the leading candidate 
of the new physics beyond the SM. We must impose on the MSSM a 
new parity called {\it R-parity} $({\rm R}_p = (-1)^{3B + L +2s})$ 
to conserve the baryon and lepton numbers, in contrast to the SM in 
which these quantum numbers are automatically conserved due to the
 gauge invariance. The violation of the R-parity causes  many exotic 
phenomena such as the proton decay. Although its conservation is 
generally assumed, this assumption is completely {\it ad hoc}. 
The  RPV interactions may exist which could be unveiled by 
upcoming high-precision experiments.   
Until now, many allowed regions and upper bounds have been 
established for single and combined parameters of the RPV 
 interactions by means of the weak processes
 \cite{barger,bhatt,barbier,dreiner}.

Herczeg \cite{herczeg1} analyzed the $d \rightarrow u e^- \bar{\nu}_e$
transition by including the RPV interactions and pointed out
that they  generate two types of new interactions.
One is the usual V-A interactions and the other is of the
scalar and pseudoscalar type.
It was shown that rather severe upper bounds on the
RPV interactions \cite{herczeg2} were obtained 
from the contribution of the pseudoscalar interaction
of the  pion decay $\Gamma (\pi \rightarrow e \nu_e ) /
 \Gamma (\pi \rightarrow \mu \nu_\mu )$ \cite{herczeg3}.
The bound on the scalar interaction was also derived by the use of electroweak 
radiative corrections \cite{khriplovich}
and  the atomic EDM  data \cite{herczeg2}.

Recently, new data sensitive to the
scalar interaction have been reported.
Hardy and Towner \cite{hardy}  updated their 
analysis on the superallowed Fermi transitions and gave an
 improved limit on the real part of the scalar interaction
 between the hadron and lepton currents.
 Meanwhile, the experiment at PSI \cite{kozela} has
 measured the transverse polarization of electrons emitted in 
 the neutron beta decay, and has given observables sensitive to
 both the real and imaginary parts of the scalar interaction. 

As the scalar interaction is sensitive to the RPVMSSM contributions
 at the tree level \cite{herczeg1}, 
the present analysis of neutron beta decay will provide useful
 constraints on RPV
interactions complementary to those obtained from the pseudoscalar
 interaction of the pion decay and
the atomic EDM via one-loop corrections.

The purpose of the present paper is to
 perform the quantitative analysis of the RPV contributions 
to the neutron beta decay in the 
light of \cite{hardy} and \cite{kozela}.
In order to explore a possibility to
investigate the scalar interaction from the neutron beta decay, 
we use the formula of angular correlation coefficients given in 
literatures \cite{jackson,ebel}. We notice here assumptions we made 
in this paper, which 
are the followings: (a) The R-parity conserving sfermion 
sector of the MSSM is assumed to be flavour diagonal and with no CP phases.
 (b) The  RPV  sector of the MSSM does not contain any 
soft SUSY breaking terms.

This paper is structured as follows. In sec. \ref{sec:RPV}  
we first construct the effective interaction of the neutron 
beta decay within RPVMSSM. 
In sec. \ref{sec:ang_corr}, we explore the angular correlation coefficients
by including neutrino momentum in the electron polarization term.
In sec. \ref{sec:analysis}, we obtain new constraints on the 
 RPV couplings from the recent data of nuclear beta decay.
Sec. \ref{sec:summary} is devoted to summary.

\section{\label{sec:RPV}Effective interaction from R-parity violation}
\subsection{\label{sec:RPVlag}R-parity violating Lagrangian}

The first step of the estimation is to construct 
 the tree level amplitude of the quark beta decay. 
The relevant interactions can be obtained from the RPV 
part of the superpotential which can be written as follows:
\begin{equation}
\hskip-1cm
W_{{\rm R}\hspace{-.5em}/} = \frac{1}{2} \lambda_{ijk} \epsilon_{ab}
 L_i^a L_j^b (E^c)_k
 +\lambda'_{ijk} \epsilon_{ab} L_i^a Q_j^b ( D^c)_k 
+ \frac{1}{2} \lambda''_{ijk} \epsilon_{lmn} (U^c)_i^l (D^c)_j^m (D^c)_k^n \ ,
\label{eq:superpotential}
\end{equation}
with $i,j,k=1,2,3$ indicating the generation, $a,b=1,2$ the $SU(2)$
indices, $l,m,n=1,2,3$ the colour indices. 
$L$ and $E^c$ denote the lepton doublet and singlet left-chiral 
superfields. $Q$, $U^c$ and $D^c$ denote the quark doublet, 
up quark singlet and down quark singlet left-chiral superfields,
 respectively. The third term in (\ref{eq:superpotential}) is 
baryon number violating, 
 so is not relevant to our subsequent discussions. We therefore 
disregard it hereafter.
By taking the F-terms of $W_{{\rm R}\hspace{-.5em}/}$, we obtain  
lepton number violating Yukawa interactions and also scalar four-point 
interactions. In our discussion, the scalar four-point interactions are 
irrelevant since we cannot construct diagrams which involve them at the 
tree level. The RPV interactions contributing to the neutron beta 
decay is as follows:

\begin{eqnarray}
{\cal L} &=&-\frac{1}{2}\sum_{ijk} \lambda_{ijk} \Bigl\{
   \tilde e_{Rk}^\dagger \bar \nu^c_{i} P_L e_j
 + \tilde e_{Lj} \bar e_k P_L \nu_i 
 + \tilde \nu_i \bar e_k P_L e_j \nonumber\\
&&\ \ \ \ \ \ \ \ \ \ \ \ \ \ \ 
 - \tilde e_{Rk}^\dagger \bar e_i^c P_L \nu_j 
 - \tilde e_{Li} \bar e_k P_L \nu_j 
 - \tilde \nu_j \bar e_k P_L e_i
                                \Bigr\} \nonumber\\
&&-\sum_{ijk} \lambda '_{ijk} \Bigl\{
   \tilde d^\dagger_{Rk} \bar \nu^c_i P_L d_j 
 + \tilde d_{Lj} \bar d_k P_L \nu_i 
 + \tilde \nu_i \bar d_k P_L d_j \nonumber\\
&&\ \ \ \ \ \ \ \ \ \ \ \ \ \ \  
 - \tilde d^\dagger_{Rk} \bar e^c_i P_L u_j 
 - \tilde e_{Li} \bar d_k P_L u_j 
 - \tilde u_{Lj} \bar d_k P_L e_i
                           \Bigr\} + \ {\rm h.c.} \ , 
\label{eq:lagrangian}
\end{eqnarray}
with $P_L = \frac{1}{2} (1-\gamma_5)$.

\subsection{\label{sec:feynman}Feynman diagrams}
\begin{figure*}[htb]
\begin{indented}
\item[]
\includegraphics[width=12cm]{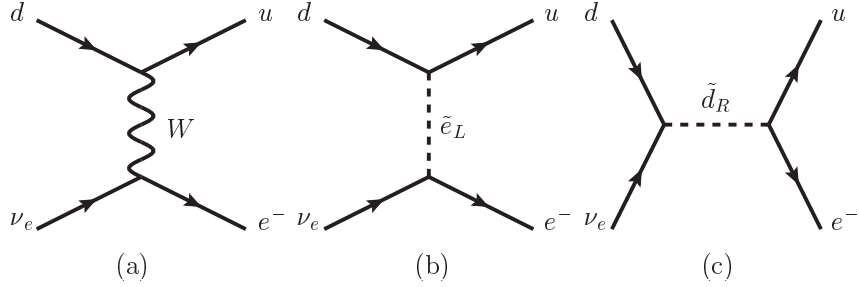}
\end{indented}
\caption{\label{fig:diagrams}Tree level amplitude of the 
quark beta decay within RPVMSSM.}
\end{figure*}

We can now build the amplitude of the quark beta 
decay within RPVMSSM at the tree level. The Feynman 
diagrams are shown in Fig.~\ref{fig:diagrams}. 

Fig.~\ref{fig:diagrams} (a)  is the  SM contribution at the lowest order. 
Its expression is as follows:
\begin{equation}
{\cal M}_{\rm SM} =
\frac{G_F}{\sqrt{2}} V_{ud} \ \bar u \gamma^\mu (1 - \gamma_5 )  d
\cdot \bar e \gamma_\mu (1 - \gamma_5 )  \nu_e 
\ , 
\end{equation}
where $G_F 
=(1.16637 \pm 0.000 01) \times 10^{-5} {\rm
 GeV}^{-2}$ \cite{ritbergen} is the Fermi constant and 
$V_{ud}=0.97418 \pm 0.00027$ \cite{savard} is the CKM matrix element.

In the tree approximation, the RPV contributions to the
$d \rightarrow u + e^- + \bar{\nu}_e$ process
are the selectron($\tilde e_L$)  (Fig.~1 (b))
and the squark($\tilde d_R$)  (Fig.~1 (c)) 
exchange  mechanisms \cite{herczeg1}.
The amplitude of the
selectron exchange between the quark and lepton
in Fig.~\ref{fig:diagrams} (b) can be written as

\begin{eqnarray}
{\cal M}_{\tilde e_L}&=&
\sum_{i=2,3} \frac{(\lambda_{1i1}-\lambda_{i11}) 
\lambda '^*_{i11} }{8 m^2_{\tilde e_{Li}}}
\bar u  \left ( 1 + \gamma_5 \right ) d \cdot
\bar e \left ( 1 - \gamma_5  \right ) \nu_e \nonumber\\
&=&\sum_{i=2,3} \frac{\lambda_{1i1} \lambda '^*_{i11} }{4 m^2_{\tilde e_{Li}}}
\bar u (1+\gamma_5 )d \cdot
\bar e (1-\gamma_5 ) \nu_e \ , 
\label{eq:mel}
\end{eqnarray}
where we have used the antisymmetry of the couplings $\lambda_{ijk}$ under 
the interchange of the first and second index
($\lambda_{ijk}=-\lambda_{jik}$). It is interesting  to note 
that ${\cal M}_{\tilde e_L}$ 
is of the type of the scalar interaction.

The  RPV couplings are constrained by extensive analyses 
on current experimental data. 
The upper limits of the  RPV couplings used in (\ref{eq:mel}) are 
given by the  first four entries in Table \ref{table:lambda}. 
The upper limits of $\lambda_{1i1},\lambda'_{11i}$ for $i=2,3$
are obtained from charged current (CC) universality and decay 
ratios $\Gamma (\tau \rightarrow e \nu \bar \nu) / \Gamma 
(\tau \rightarrow \mu \nu \bar \nu)$,
 $\Gamma (\pi \rightarrow e \nu) / \Gamma (\pi \rightarrow \mu \nu)$
 and $\Gamma (\tau \rightarrow \pi \nu_\tau) / 
\Gamma (\pi \rightarrow \mu \nu)$ \cite{barger, bhatt,barbier}. 
As we see,  the magnitude of ${\cal M}_{\tilde e_L}$ is small 
compared with that of ${\cal M}_{\rm SM}$. This scalar interaction is, however, 
 potentially 
important since it could give rise to observables in  neutron beta decay 
to which $V-A$ interactions do not contribute.

\begin{table}
\caption{\label{tab:upper_limit} 
Current upper limits of (the magnitude of) RPV  couplings \cite{tata,faessler}. 
Expressions with square brackets correspond to their masses in unit of 100 GeV. 
$\tilde  q$ and $\tilde g$ refer to the squark and gluino, respectively.
}
\begin{indented}
\item[]
\begin{tabular}{cl}
\br
Coupling constants & Upper bounds \\
\mr
$\vert \lambda_{121}\vert $ & $< 0.049 \ [m_{\tilde e_R}]$\\
$\vert \lambda_{131}\vert $ & $< 0.062\ [m_{\tilde e_R}]$\\
$\vert \lambda'_{211} \vert $ & $< 0.059\ [m_{\tilde d_R}]$\\
$\vert \lambda'_{311} \vert $ & $< 0.11 \ [m_{\tilde d_R}]$\\
$\vert \lambda'_{111}\vert $ & $< 1.3 \times 10^{-4} \cdot
 [m_{\tilde q}]^2 [m_{\tilde g}]^{1/2} $\\
$\vert \lambda'_{112}\vert $ & $< 0.021 \ [m_{\tilde s_R}]$\\
$\vert \lambda'_{113}\vert $ & $< 0.021 \ [m_{\tilde b_R}]$\\
\br
\end{tabular}
\end{indented}
\label{table:lambda}
\end{table}

The squark exchange amplitude in Fig.~\ref{fig:diagrams} (c) 
can be written as
\begin{eqnarray}
{\cal M}_{\tilde d_R}&=&
\sum_{i=1,2,3} \frac{|\lambda'_{11i} |^2}{4m^2_{\tilde d_{Ri}}}
\bar \nu_e^c ( 1 - \gamma_5  ) d \cdot
\bar u ( 1 + \gamma_5 ) e^c \nonumber\\
&=&\sum_{i=1,2,3} \frac{|\lambda'_{11i} |^2}{8m^2_{\tilde d_{Ri}}}
\bar u \gamma^\mu (1 - \gamma_5 ) d
\cdot
\bar \nu_e^c \gamma_\mu (1 + \gamma_5 ) e^c  \nonumber\\
&=& \sum_{i=1,2,3} \frac{|\lambda'_{11i} |^2}{8m^2_{\tilde d_{Ri}}}
\bar u \gamma^\mu (1 - \gamma_5 ) d
\cdot
\bar e \gamma_\mu (1 - \gamma_5 ) \nu_e   \ .
\label{eq:mdr}
\end{eqnarray}
To obtain the final result in (\ref{eq:mdr}), we have used the Fierz transformation. 
(\ref{eq:mdr}) has  the $V-A$ type interaction like the SM.  
Here the constraints on the RPV couplings appearing  
in (\ref{eq:mdr}) are given by the last three entries in 
Table \ref{table:lambda}. 
The couplings $\lambda'_{i11}$ ($i=1,2,3$) are constrained from the analysis of the 
double beta decay \cite{faessler} and CC universality \cite{barger, bhatt, barbier}.
In our analysis, we use the empirical value of $V_{ud}$ and axial vector
 coupling constant $g_A$, therefore the effect of (\ref{eq:mdr}) is 
absorbed into those coupling constants.

\subsection{\label{sec:effint}Effective interaction}
The next step is to build the effective interaction of the neutron 
beta decay process. To obtain it, we must construct the nucleon matrix
elements 
from the quark beta decay amplitudes. The effective 
interaction can be written as
\begin{equation}
H_\beta = H_{\rm SM} +H_{{\rm R}\hspace{-.5em}/} \ ,
\end{equation}
where the first term is the SM contribution and the second term the
contribution from  Fig.~\ref{fig:diagrams} (b). The nucleon matrix
elements are described by form factors  as follows:
\label{eq:nucl_mat_elem}
\begin{eqnarray}
\langle p | \bar u \gamma^\mu d | n \rangle &=& g_V (q^2) 
\bar p \gamma^\mu n \ , \\
\langle p | \bar u \gamma^\mu \gamma_5 d | n \rangle &=&
 g_A (q^2) \bar p \gamma^\mu \gamma_5 n \ , \\
\langle p | \bar u d | n \rangle &=& g_S (q^2) \bar p n \ ,
\label{eq:gs}
\end{eqnarray}
where we have explicitly written the nucleon fields 
as $n$ and $p$, and $q$ is the momentum transfer. The  form factor of 
the pseudoscalar interaction need not be introduced 
 since its contribution  vanishes in the non-relativistic limit. The 
induced form factors were also neglected. In this paper, we use the
 constants $g_V=g_V (0)$, $g_A=g_A (0)$, $g_S=g_S (0)$ since $q^2 $ is
 much smaller than the nucleon mass. 
Here recall that $g_V=1$ from the CVC and $g_A=1.2739 \pm 0.0019$ from
 \cite{abele}.

The determination of $g_S$ needs more involved discussion. Within 
approximate isospin symmetry, the matrix element
 $\langle p | \bar u d | n \rangle = \langle p | \bar u u -
 \bar d d | p \rangle$. 
According to Gasser and Leutwyler \cite{gasser}, the isospin 
asymmetric part of the nucleon form factor can be written as
\begin{equation}
\delta = \frac{m_d - m_u }{2M_N} \langle p | \bar u u - \bar d d | p \rangle \ ,
\label{eq:delta1}
\end{equation}
where $M_N$ indicates the nucleon mass. To the lowest order 
expansion in up and down quark mass difference, the neutron 
and proton mass difference equals  $\delta$:
\begin{equation}
\delta = M_n -M_p = (2.05 \pm 0.30) \ {\rm MeV} \ ,
\label{eq:delta2}
\end{equation}
where the mass difference $M_n - M_p$ are of pure QCD origin, i.e., 
without electromagnetic contribution. From (\ref{eq:gs}), (\ref{eq:delta1}) and 
(\ref{eq:delta2}) we obtain
\begin{equation}
g_S = \frac{M_n-M_p}{m_d - m_u} = 0.49 \pm 0.17 \ ,
\label{eq:numericalgs}
\end{equation}
where we have used $m_d =9.3 \pm 0.9$ MeV, 
$m_u / m_d =0.553 \pm 0.043 $ \cite{leutwyler}. 
The error  in (\ref{eq:numericalgs}) is  
mostly due to the ambiguities of the current quark mass. 
Note that the  values of $g_S$ obtained previously by 
Adler {\it et al.}\cite{adler},  and by Wakamatsu \cite{wakamatsu}
 are consistent with (\ref{eq:numericalgs}).

We can now construct the effective interaction of the neutron
 beta decay within RPVMSSM at the tree level, which is given by
\begin{equation}
H_{\rm SM} = \frac{G_F}{\scriptstyle \sqrt{2}} V_{ud} \, \bar p
 \gamma^\mu (g_V -g_A \gamma_5 ) n \cdot \bar e \gamma_\mu (1-\gamma_5
 )\nu_e
 \label{eq:hsm}
\end{equation}
and
\begin{equation}
H_{{\rm R}\hspace{-.5em}/} = C_S\, \bar p  n \cdot \bar e (1-\gamma_5 )\nu_e + {\rm h.c.}
\label{eq:hrpv}
\end{equation}
where
\begin{equation}
C_S = g_S \sum_{i=2,3} \frac{\lambda_{1i1} \lambda '^*_{i11} }{4
 m^2_{\tilde e_{Li}}} \ . 
\label{eq:cscp}
\end{equation}

\section{\label{sec:ang_corr}Angular correlations of the neutron beta
 decay}

The formula for the  angular correlation coefficients of the 
beta decay by taking into
account of  the  polarized neutron and electron
is given by  Jackson, Treiman and Wyld (JTW) \cite{jackson},
and extended by Ebel and Feldman \cite{ebel} as follows:
\begin{eqnarray}
&&\omega (E_e , \Omega_e , \Omega_{\nu})\propto 
1 + a \frac{{\vec p_e} \cdot {\vec p_{\nu}}}{E_e E_{\nu}}
+b \frac{m_e}{E_e} \nonumber \\
&&\ \ \ + \ \vec \sigma_n \ \cdot 
\left\{
A\frac{{\vec p_e}}{E_e} + B\frac{{\vec p_{\nu}}}{E_{\nu}}
+D\frac{{\vec p_e}\times {\vec p_{\nu}}}{E_e E_{\nu}} 
\right\} \nonumber\\
&&\ \ \ +\ \vec \sigma_e \ \cdot 
\left\{ 
G \frac{\vec p_e }{E_e}+H
 \frac{\vec p_{\nu}}{E_{\nu}}
+K\frac{\vec p_e}{E_e +m_e}\frac{\vec p_e \cdot \vec p_{\nu}}{E_e E_{\nu}}
+L\frac{\vec p_e \times \vec p_{\nu}}{E_e E_{\nu}} 
\right\} \nonumber\\
&&\ \ \ +\ \vec \sigma_e \ \cdot 
\left\{
 N \vec \sigma_n
+Q \frac{\vec p_e}{E_e +m} \frac{ \vec \sigma_n \cdot \vec p_e}{E_e}
+ R \frac{ \vec \sigma_n \times \vec p_e}{E_e} 
\right\} \nonumber\\
&&\ \ \ + \ \vec \sigma_e \ \cdot 
\left\{
 S \vec \sigma_n \frac{\vec p_e \cdot \vec p_\nu}{E_e E_\nu}
+T \vec p_e \frac{\vec \sigma_n \cdot \vec p_\nu}{E_e E_\nu}
+U \vec p_\nu \frac{\vec \sigma_n \cdot \vec p_e}{E_e E_\nu}
+V \frac{\vec p_\nu \times \vec \sigma_n}{E_\nu} \right.
\nonumber\\
&&\left. \ \ \ \ \ \ \ \ \ \ \ \ \ \ \ \ \ + \ W \frac{\vec p_e }
{E_e + m_e} \frac{\vec \sigma_n \cdot 
\vec p_e \times \vec p_\nu}{E_e E_\nu}
\right\} . 
\label{eq:decay_distribution}
\end{eqnarray}

Here $\vec p_e $ and $\vec p_{\nu}$ are respectively the momenta of the
electron and the neutrino, $\vec \sigma_n $ the neutron spin
polarization and $\vec \sigma_e$ the final state electron polarization. 
The angular correlations of the last line (terms with $S$, $T$, $U$,
$V$ and $W$ coefficients) are correlations with momentum of emitted
anti-neutrino, polarizations of initial neutron and emitted electron, 
given by Ebel and Feldman \cite{ebel}.

The angular correlation coefficients are evaluated using the main term of
vector and axial vector current of the SM  in (\ref{eq:hsm}) and
the scalar  interaction in (\ref{eq:hrpv}).
Here we take static approximation of nucleon current and keep only the
leading order contributions.  The results are summarized in Table 
\ref{table:coff}
where we used

\begin{equation}
\lambda = \frac{g_A}{g_V }, 
\end{equation}

\begin{equation}
\alpha_{R} = \frac{ 2  Re (C_S)}{\frac{G_F}{\sqrt{2}}
V_{ud}g_V(1 + 3\lambda^2)}, 
\end{equation}
\begin{equation}
\alpha_{I} = \frac{ 2  Im (C_S)}{\frac{G_F}{\sqrt{2}}
V_{ud}g_V(1 + 3\lambda^2)}.
\end{equation}

There are quite a few terms which are sensitive to the scalar
interaction (see Table \ref{table:coff}).
The coefficients $b,S,U$ and $R,L,V,W$ probe respectively 
 the real and the imaginary part of the scalar interaction.
Among them, $S,T,V$ and $W$, in addition to the $b,R$ and $L$ 
which already studied by JTW, are particularly interesting observables.
Since the main term of the SM (or $V-A$) does not contribute to those
observables, there is a possibility to
probe new physics through the scalar interaction. 
The effects of the final state interaction (FSI) have been 
in \cite{jackson,vogel,ando}.

It is also noticed that the $B$ correlation
can be used to investigate the scalar interaction.
Though  $V-A$ interactions contribute to $B$,
one may be able to extract a small effect of the scalar
interaction  by using 
the extra energy dependence  $\frac{m_e}{E_e}$ in the scalar 
contribution \cite{ramsey}.

\begin{table}
\caption{Angular correlation coefficients of the neutron beta decay}
\begin{indented}
\item[]
\begin{tabular}{ccc}
\br
Coefficients & SM & RPV \\
\mr
$a$  & $(1 - \lambda^2)/(1 + 3 \lambda^2)$      & $0$     \\
$b$  & $0$                     & $\alpha_R$                \\
$A$ & $2\lambda(1 - \lambda)/(1+3\lambda^2)$    & $0$       \\
$B$ & $2\lambda(1+\lambda) / (1 + 3\lambda^2)$ & $\lambda\alpha_R (m_e / E_e)$\\
$D$ & $0$                       &  $0$                  \\
$G$ &  $-1$                     &  $0$                   \\
$H$ & $(m_e / E_e)(\lambda^2-1) /(1 + 3 \lambda^2)$    & $- \alpha_R$    \\
$K$ & $(\lambda^2-1)/(1 + 3 \lambda^2)$     & $ \alpha_R$     \\
$L$ & $0$                                   & $  \alpha_I$    \\
$N$ & $-(m_e / E_e)2\lambda(1-\lambda)/(1+3\lambda^2)$ &
	 $-\lambda\alpha_R$
      \\
$Q$ & $-2\lambda(1-\lambda)/(1+3\lambda^2)$  & $\lambda\alpha_R$ \\
$R$ & $0$                       & $-\lambda\alpha_I$ \\
$S$ & $0$                       & $\lambda\alpha_R$ \\
$T$ & $-2\lambda(1+\lambda)/(1 + 3\lambda^2)$ &  $0$ \\
$U$ & $0$                      & $-\lambda\alpha_R$ \\
$V$ & $0$                       & $-\lambda\alpha_I$ \\
$W$& $0$                       & $\lambda\alpha_I$ \\
\br
\end{tabular}
\end{indented}
\label{table:coff}
\end{table}

\section{\label{sec:analysis}Analyses}
As we have seen, the RPV interaction of (\ref{eq:lagrangian}) 
generates (\ref{eq:mel}) which turns out to be scalar 
interactions of the neutron beta decay (\ref{eq:hrpv}). 
Combining the current upper limits of Table \ref{table:lambda}, 
we have the following bounds for each combination of RPV couplings
 in (\ref{eq:cscp}):
\begin{equation}
\vert \lambda_{121} \lambda'^*_{211} \vert < 0.0029 
[m_{\tilde e_R }] [m_{\tilde d_R}]\ ,
\label{eq:17a}
\end{equation}
\begin{equation}
\vert \lambda_{131} \lambda'^*_{311} \vert < 0.0068
 [m_{\tilde e_R }] [m_{\tilde d_R}] \ ,
\label{eq:17b}
\end{equation}
where $[m_{\rm SUSY}] = m_{\rm SUSY} / (100\, {\rm GeV})$.
 (\ref{eq:17a}) and  (\ref{eq:17b}) will be compared with 
the analysis confronted with \cite{hardy} and \cite{kozela}. 

From the recent update of the analysis on 20 superallowed 
Fermi transitions, Hardy and Towner \cite{hardy} have given 
a new bound on the Fierz interference term $b_F$ 
which corresponds to $b$ term in (\ref{eq:decay_distribution}). 
This one is a probe of the real part of the scalar interaction.
By fitting $\overline{{\cal F}t}$ values of the superallowed 
Fermi transitions, they obtained 
\begin{equation}
\frac{b_F}{2}=\frac{Re(C_S)}{C_V} = +0.0011\pm 0.0013 \ ,
\end{equation}
where $C_V = V_{ud} \frac{G_F}{\sqrt 2} g_V$.
From this, we obtain the following limit on the 
combined RPV couplings:
\begin{equation}
Re \Bigl( \sum_{i=2,3} \lambda_{1i1} \lambda'^*_{i11} \Bigr)
 = (7.2 \pm 8.5 ) \times 10^{-4} \frac{[m_{\tilde e_L}]^2}{[g_S]} \ ,
\label{eq:relim}
\end{equation}
where $[g_S]\equiv \frac{g_S }{0.49}$. 

The complementary information on the imaginary part
of the RPV couplings 
can be obtained from the $R$ correlation of the neutron 
beta decay distribution. The angular correlation $R$,
which is the triple product of the polarization and the momentum 
of emitted electron, and polarization of the initial neutron,
 has been  recently measured at PSI \cite{kozela} as 
\begin{equation}
R= 0.008 \pm 0.011 \pm 0.005 .
\label{eq:RPSI}
\end{equation} 
The contribution of FSI to $R$ can be estimated as
\begin{equation}
R_{\rm FSI} =8.57 \times 10^{-4} \times \frac{m_e}{p_e} \ ,
\end{equation}
according to the formula of JTW \cite{jackson}.
Here $m_e$ and $p_e$ are respectively the mass and 
the magnitude of the spatial momentum of the emitted electron. 
The effect of FSI is by an order of magnitude smaller 
than the experimental data (\ref{eq:RPSI}).
Using the experimental value (\ref{eq:RPSI}) neglecting FSI,
we obtained the following constraint on the imaginary part of the
coupling constants,
\begin{equation}
Im \left( \sum_{i=2,3} \lambda_{1i1} \lambda'^*_{i11} \right) 
=( -0.012 \pm 0.017 \pm 0.008 ) \frac{[m_{\tilde e_L}]^2}{[g_S]} \ .
\label{eq:imlim}
\end{equation}

\begin{figure}[h]
\begin{indented}
\item[]
\begin{center}
\includegraphics[width=8cm]{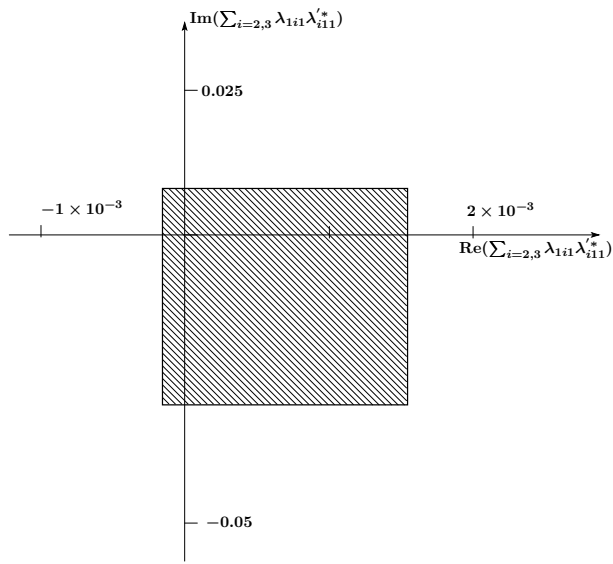}
\end{center}
\caption{Allowed region of $\sum_{i=2,3} \lambda_{1i1}
 \lambda'^*_{i11}$.
 We have set $m_{\tilde e_L}=100\, {\rm GeV}$ and $g_S = 0.49$.}
\end{indented}
\end{figure}

The obtained limits on the RPV couplings in this analysis
are summarized in Fig. 2. Here we have taken all s-particle masses to
100 GeV. As we can see, a strong constraint of the
real part of the RPV parameters is obtained from $b$.
On the other hand, the constraint of the imaginary part of the 
RPV parameters might be comparable to  (\ref{eq:17a}) and  (\ref{eq:17b}).
However, the limits obtained in (\ref{eq:relim}) and (\ref{eq:imlim}) 
depend on the mass $m_{\tilde e_L}$ while 
the current ones ( (\ref{eq:17a}) and  (\ref{eq:17b})) 
depends on the masses
$m_{\tilde e_R}$ and $m_{\tilde d_R}$.
Therefore  (\ref{eq:relim}) and (\ref{eq:imlim}) give 
a new type of constraint.
With the progress of experiment on $R$,
one will be able to further constrain the imaginary part of 
the coupling constant.

It is of interest to compare our constraints, (\ref{eq:relim}) 
and (\ref{eq:imlim})
with those obtained previously. It has been pointed out in \cite{khriplovich}
that usual electroweak radiative corrections could transform the T-odd,
P-even interactions into a T-odd, P-odd ones, and that one could
put limits on the T-odd, P-even interactions by using the rich data
on the part of T-odd, P-odd interactions.
Herczeg \cite{herczeg2} considered such radiative corrections in RPV
models and confronted the induced interactions with EDM data.
The analysis depends on the sparticle masses, but in general
more severe constraints can be made available than (\ref{eq:imlim}).

It has been pointed out in \cite{herczeg3} that the ratio
$\Gamma(\pi \rightarrow e \nu)/\Gamma(\pi \rightarrow \mu \nu)$
is a sensitive proble to the pseudoscalar type $\bar{d}u \rightarrow e
\nu_e$ interactions. For the case of the RPV contribution (\ref{eq:mel}),
The scalar and the pseudoscalar interactions are of the
same strength, and one can deduce an upper bound on $C_S/g_S$ by
employing the data of the ratio. The upper bound
obtained in such a manner is a fraction of the value (\ref{eq:relim}).

\section{\label{sec:summary}Summary}

To summarize, we have studied the contribution of 
the RPVMSSM to the angular correlation 
coefficients and the Fierz interference term in the neutron beta decay. 
We have deduced  new bounds on the real and imaginary parts of the 
combinations of RPV couplings as in (\ref{eq:relim}) and (\ref{eq:imlim}) through the scalar interaction (\ref{eq:hrpv}) of the beta decay.
These are based on the recent data of \cite{hardy} 
and \cite{kozela} respectively.
The obtained upper bound for $Re(\lambda_{1j1}\lambda'^*_{j11})$
is consistent with a stronger constraint
obtained from the analysis of pion decay \cite{herczeg1,herczeg2,herczeg3}.
With the hypothesis of the single coupling dominance, 
(\ref{eq:relim}) and (\ref{eq:imlim}) can be replaced by a constraint 
for either $ \lambda_{121} \lambda'^*_{211}$ or $ \lambda_{131} 
\lambda'^*_{311}$.

We have also discussed 
angular correlations $S,T ,U,V$ and $W$, 
which involve emitted electron polarization and neutrino momentum. In addition 
to the known ones, coefficients $b,S,U$ and $R,L,V,W$ are respectively
 sensitive to the real and imaginary parts of the scalar interaction.
Their precise measurements are therefore of great interest for further limiting
the RPV couplings. 
Regarding the CP violating (imaginary) ones, the precise evaluation of 
the FSI contribution is also needed, as done in \cite{vogel} or \cite{ando}.

\ack

The authors thank Prof. M. Wakamatsu for illuminating discussions.
 This work is supported by the Japan Society for the Promotion 
of Science, Grant-in-Aid for Scientific Research(C) 20540270.

\end{document}